\documentclass[11pt]{article}
\pdfoutput=1 

\usepackage[margin=1in]{geometry}
\usepackage{graphicx}
\usepackage{amsfonts}
\usepackage{booktabs}

\usepackage{amssymb,amsmath,amsthm,mathtools,thmtools,xspace}
\usepackage{MnSymbol,wasysym}
\usepackage{enumitem}

\usepackage{bussproofs}

\usepackage{hyperref}
\usepackage[usenames,dvipsnames]{xcolor}
\hypersetup{
	colorlinks=true,
	linkcolor=Sepia,
	citecolor=Sepia,
	filecolor=Sepia,
	urlcolor=Sepia
}

\usepackage[english]{babel}

\newcommand{\B}{\ensuremath{\{0,1\}}}
\newcommand{\Search}{\ensuremath{\mathsf{Search}}}
\newcommand{\cert}{\ensuremath{\mathsf{cert}}}
\newcommand{\set}[1]{\ensuremath{\{#1\}}}
\newcommand{\KW}{\ensuremath{\mathsf{KW}}}
\newcommand{\mKW}{\ensuremath{\mathsf{mKW}}}
\newcommand{\IND}{\ensuremath{\text{\scshape Ind}}}

\newcommand{\stconn}{\ensuremath{\text{\scshape{stConn}}}}
\newcommand{\clique}{\ensuremath{\text{\scshape{Clique}}}}
\newcommand{\cliquecol}{\ensuremath{\text{\scshape{Clique\text{-}Col}}}}
\newcommand{\gen}{\ensuremath{\text{\scshape{Gen}}}}
\newcommand{\xorsat}{\ensuremath{\text{\scshape{Xor\text{-}Sat}}}}
\newcommand{\xor}{\ensuremath{\text{\scshape{Xor}}}}

\newcommand{\newclass}[2]{\newcommand{#1}{{\text{\upshape\sffamily #2}}\xspace}}
\renewcommand{\P}{{\text{\upshape\sffamily P}}\xspace}
\newclass{\NP}{NP}
\newclass{\cc}{cc}
\newclass{\dt}{dt}
\newclass{\NC}{NC}
\newclass{\BQP}{BQP}
\newclass{\coNP}{coNP}
\newclass{\TFNP}{TFNP}
\newclass{\M}{M}
\newclass{\NL}{NL}
\newclass{\FP}{FP}
\newclass{\CLS}{CLS}
\newclass{\EML}{EOML}
\newclass{\SML}{SOML}
\newclass{\PPA}{PPA}
\newclass{\PPAD}{PPAD}
\newclass{\PPADS}{PPADS}
\newclass{\UEOPL}{UEOPL}
\newclass{\myS}{S}
\newclass{\PPP}{PPP}
\newclass{\PLS}{PLS}
\newclass{\Ppoly}{P/poly}
\newclass{\mNC}{mon-NC}
\newclass{\mAC}{mon-AC}
\newclass{\mP}{mon-P}

\DeclareMathOperator{\CP}{CP}
\DeclareMathOperator{\Res}{R}
\DeclareMathOperator{\lin}{LIN}

\renewcommand{\epsilon}{\varepsilon}

\usepackage{tikz}
\definecolor{myGold}{RGB}{231,141,20}
\definecolor{myBlue}{rgb}{0.19,0.41,.65}
\definecolor{myPurple}{RGB}{175,0,124}
\usetikzlibrary{arrows.meta,positioning} 

\addto\extrasenglish{}
\addto\extrasenglish{}

\declaretheorem[name=Theorem, parent=section]{theorem}

\declaretheorem[name=Definition, sibling=theorem, style=definition]{definition}
\declaretheorem[name=Example, numbered=no, style=definition]{example}

\declaretheorem[name=Corollary,sibling=theorem]{corollary}
\declaretheorem[name=Open problem]{problem}

\newcommand{\complclassformat}[1]%
        {\textrm{\upshape{\textsf{#1}}}}

\newcommand{\rcprotocols}{real communication protocols\xspace}
\newcommand{\rcprotocol}{real communication protocol\xspace}

\let\OLDthebibliography\thebibliography
\renewcommand\thebibliography[1]{
  \OLDthebibliography{#1}
  \setlength{\parskip}{2pt}
  \setlength{\itemsep}{1.85mm}
}

\begin{document}

\clearpage

\begin{center}
{\Large SIGACT News Complexity Theory Column, March 2022\\[2mm]
\LARGE\bf Proofs, Circuits, and Communication}\\[4mm]
%
\medskip
{\Large \bf \em
S.F. de Rezende\/}\footnote{Department of Computer Science, Lund University,
Lund, Sweden\@. {\tt susanna.rezende@cs.lth.se}. 
Supported by ELLIIT and Knut and Alice Wallenberg grant KAW 2021.0307.}
\hspace{2em}
{\Large \bf \em
M. G\"o\"os\/}\footnote{School of Computer and Communication Sciences, EPFL,
Lausanne, Switzerland\@. {\tt mika.goos@epfl.ch}.}
\hspace{2em}
{\Large \bf \em 
R. Robere\/}\footnote{School of Computer Science, McGill University,
Montreal, QC, Canada\@. {\tt robere@cs.mcgill.ca}. 
Supported by NSERC.}

\bigskip 

\includegraphics[height=37mm]{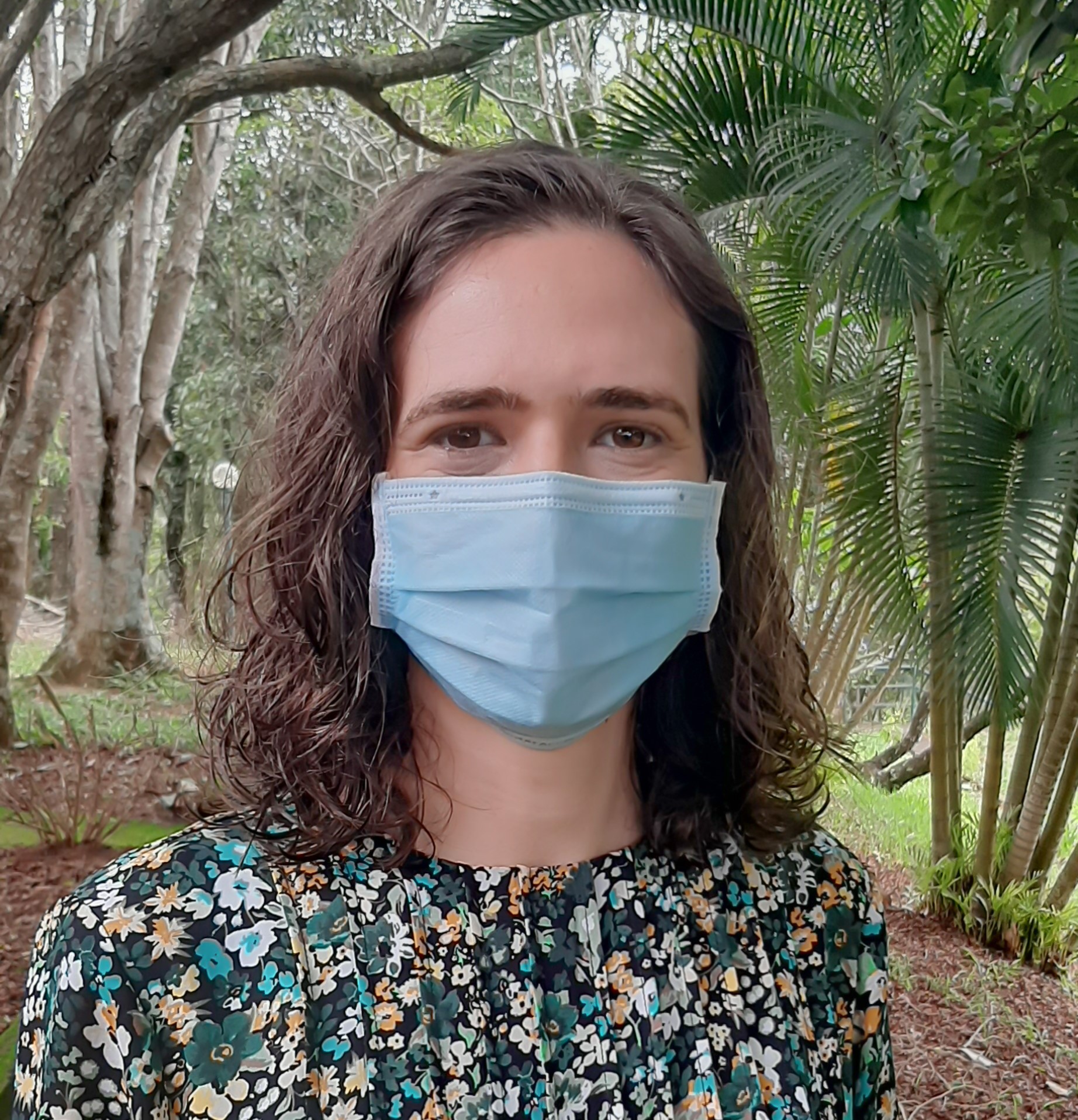}
\includegraphics[height=37mm]{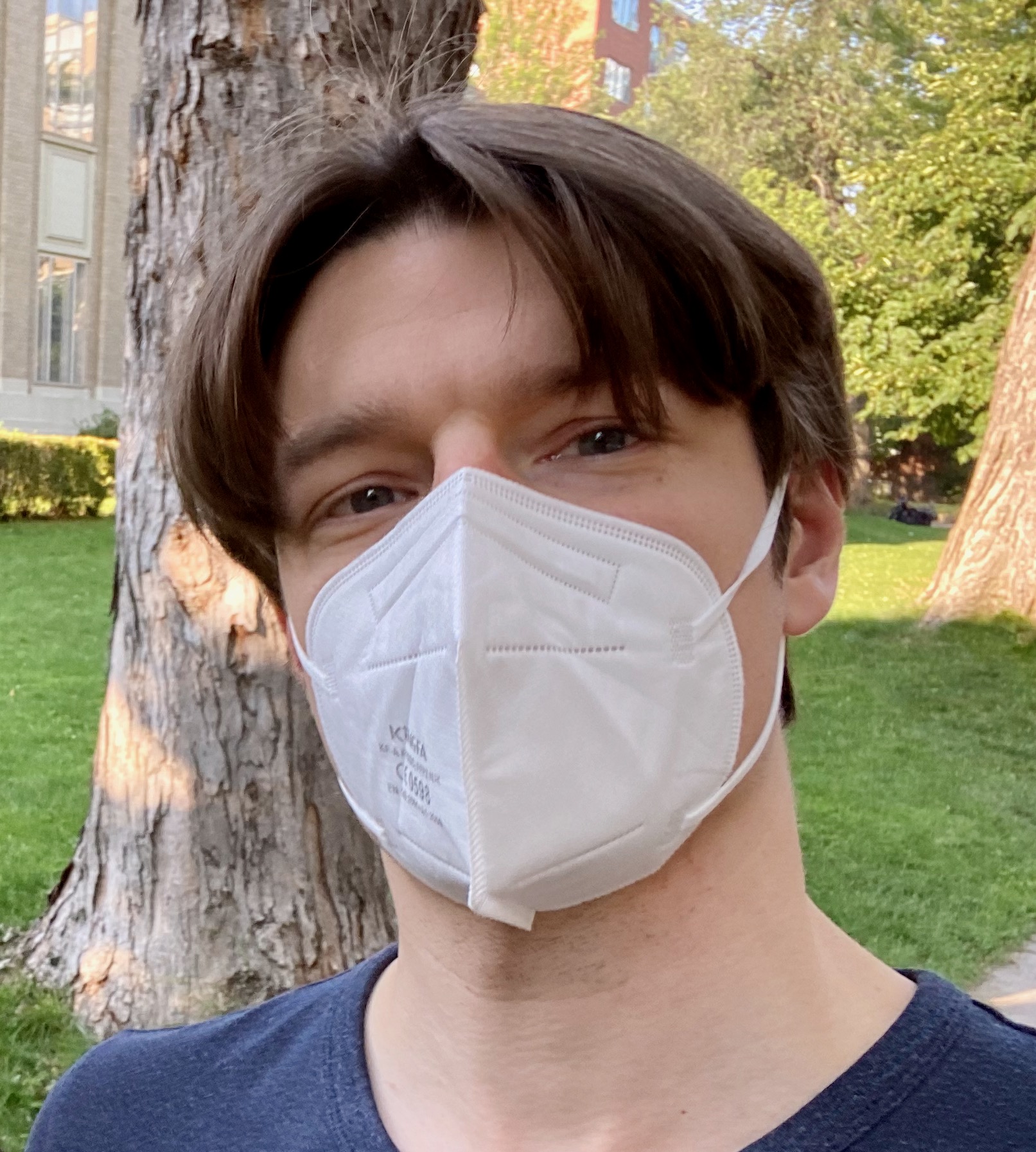}
\includegraphics[height=37mm]{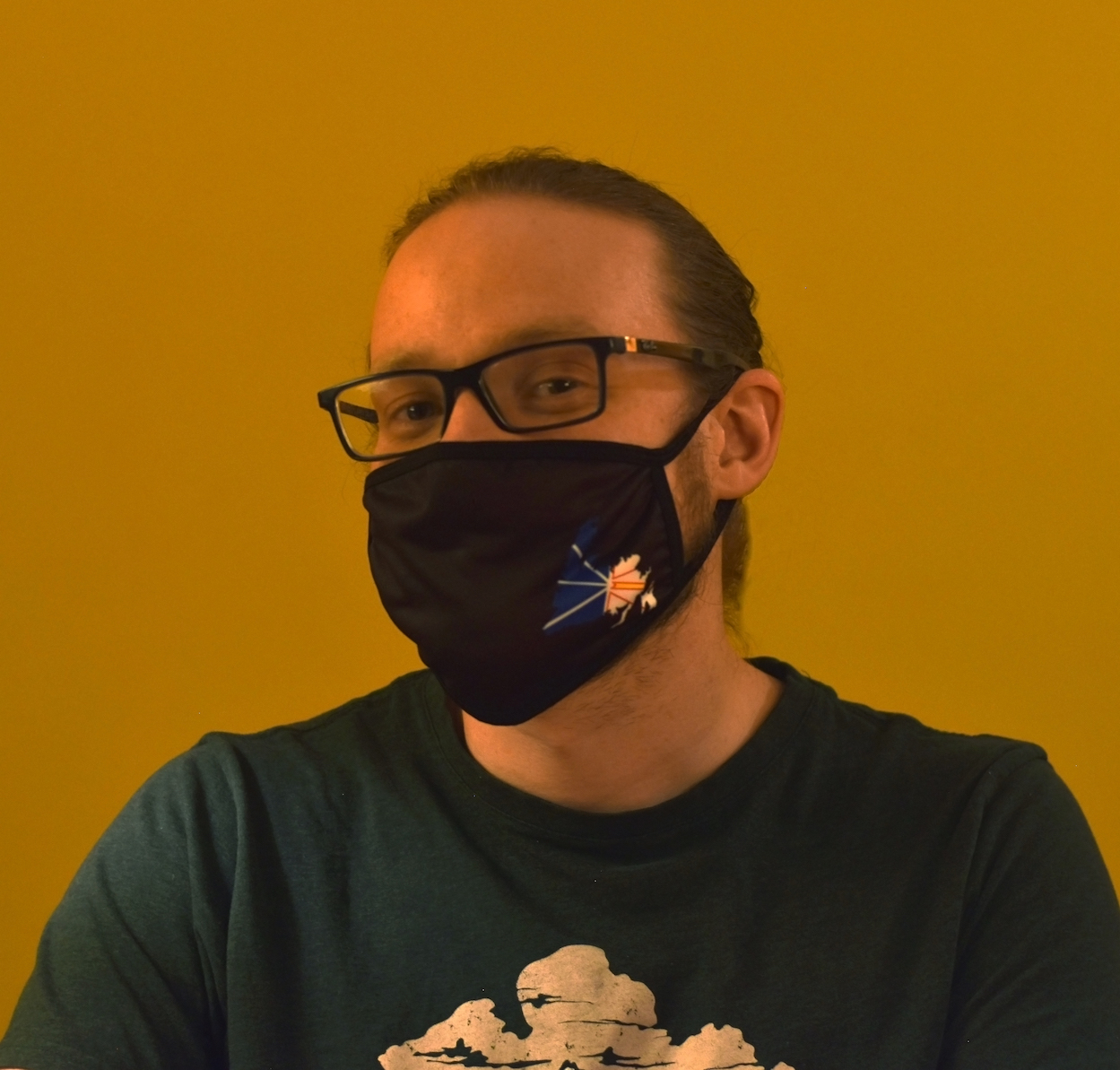}
\end{center}


\begin{abstract}
We survey lower-bound results in complexity theory that have been obtained via newfound interconnections between propositional proof complexity, boolean circuit complexity, and query/communication complexity. We advocate for the theory of \emph{total search problems} ($\TFNP$) as a unifying language for these connections and discuss how this perspective suggests a whole programme for further research.
\end{abstract}


\section{Introduction}

In recent years there has been a large number of new results in both \emph{propositional proof complexity} and \emph{boolean circuit complexity}.
These results include: optimal $2^{\Omega(n)}$ lower bounds on the size of monotone boolean formulas computing an explicit boolean function in $\NP$ \cite{PR2017}, the refinement of the $\mAC^i$ hierarchy from the $\mNC^i$ hierarchy and new tradeoffs for cutting planes proofs~\cite{dRNV2016}, a new family of techniques for proving lower bounds on cutting planes proofs and monotone circuit size \cite{GGKS2020}, and exponential lower bounds on the size of cutting planes proofs for random CNF formulas \cite{FPPR2017, HP2017}.
All of these new results have been enabled, either directly or indirectly, by two tools:
\begin{enumerate}
	\item The discovery of {\bf\itshape new connections} between proof complexity and query complexity, as well as between boolean circuit complexity and communication complexity.
	\item The development of {\bf\itshape query-to-communication lifting theorems} which show that, for certain tasks, query complexity lower bounds can be ``lifted'' to communication lower bounds.
\end{enumerate}
Taken together, these tools reveal new connections between proof complexity and monotone circuit complexity.
In particular, many standard propositional proof systems (e.g.,~resolution, cutting planes, Nullstellensatz) have been shown to have corresponding ``partner'' models of monotone computation (resp.~boolean circuits, real circuits, span programs) such that the lower bounds on proof complexity in a proof system imply lower bounds on complexity of computation in the ``partner'' model. 

A central theme in these results is the important role played by certain {\bf\itshape total search problems} in both proof complexity and boolean circuit complexity.
On the one hand, for many proof systems the complexity of refutations of an unsatisfiable CNF formula $F$ is closely related to the complexity of the so-called \emph{falsified clause search problem} $\Search(F)$, where we are given an assignment $x$ to the variables of $F$ and need to find a clause in $F$ that is false under $x$. 
Meanwhile, for boolean models of computation, the right total search problem is the famous \emph{Karchmer--Wigderson game}~\cite{Karchmer1990}, which was introduced to understand circuit complexity using communication complexity.
It has turned out to be fruitful to place these total search problems on center stage when trying to understand proof systems and circuit models. In fact, recent research has shown that the classical (Turing machine) theory of total search problems ($\TFNP$) can serve as a general ``organizing principle'' for proof systems and circuit classes in this way \cite{GKRS2019}.

\paragraph{Outline.}
This survey explores the aforementioned themes starting from base principles.
In \autoref{sec:tree-like} we will give a gentle introduction to the classical connections between \emph{tree-like resolution} and \emph{decision trees}, as well as \emph{boolean formulas} and \emph{deterministic communication protocols}.
In \autoref{sec:dags}, we will extend the results from \autoref{sec:tree-like} to dag-like models.
Then, in \autoref{sec:cutting-planes}, we describe applications of these ideas to the semi-algebraic \emph{cutting planes} proof system.
Finally, in \autoref{sec:perspectives}, we will discuss how these classical connections can be viewed as part of a more general framework, in analogy to the theory of $\TFNP$, and how this perspective suggests a programme for further research.

\section{Tree-like Resolution Proofs and Monotone Boolean Formulas}
\label{sec:tree-like}

A \emph{boolean variable} is a variable $x$ taking values in $\B$, where we interpret $0$ as ``False'' and $1$ as ``True''.
A \emph{boolean literal} is a boolean variable $x_i$ or its negation~$\overline x_i$.
A \emph{boolean formula} (also called a \emph{propositional formula} or just a \emph{formula}) is an expression composed out of boolean variables and~$\wedge, \vee$ and $\neg$.
A \emph{clause} is a disjunction of boolean literals: for example, $x_1 \vee x_2 \vee \overline x_3$ is a clause, as is $x_1$.
The \emph{empty clause}, denoted $\bot$, has no literals in it, and is always false.
A boolean formula~$F$ is in \emph{conjunctive normal form} (CNF) if it is a conjunction of clauses; for example,
\begin{equation*}
	\label{eq:ind3}
 x_1 \wedge (\overline x_1 \vee x_2) \wedge (\overline x_2 \vee x_3) \wedge \overline x_3\,.
\end{equation*}

\subsection{Resolution and decision trees}
Our starting point is a well-known connection between \emph{tree-like resolution} and \emph{decision trees}.

\begin{definition}
	Let $A, B$ be clauses of boolean literals, and let $\ell$ be any boolean literal not occurring in $A$.
	The \emph{resolution rule} is the deduction rule
	\begin{prooftree}
		\AxiomC{$A \vee \ell$}
		\AxiomC{$B \vee \overline \ell$}\RightLabel{.}
		\BinaryInfC{$A \vee B$}
	\end{prooftree}
	The \emph{weakening rule} is the deduction rule
	\begin{prooftree}
		\AxiomC{$A$}\RightLabel{.}
		\UnaryInfC{$A \vee \ell$}
	\end{prooftree}
\end{definition}

Note that the resolution rule is \emph{sound}: if $x$ is an assignment that satisfies both $A \vee \ell$ and $B \vee \overline \ell$ then $x$ also satisfies $A \vee B$.

Let $F = C_1 \wedge \cdots \wedge C_m$ be a boolean formula in CNF\@.
A \emph{resolution proof} of a clause $C$ from $F$ is a sequence of clauses \[ D_1, D_2, \ldots, D_s\,,\]
where $D_s = C$, and for each $i$ either $D_i$ is a clause from $F$ or is deduced from earlier clauses by the resolution rule or the weakening rule.
The \emph{width} of the proof is the maximum number of literals in any clause appearing in the proof.
The proof is called a \emph{refutation of $F$} if $C = \bot$, the empty clause.
It is natural to associate a proof \emph{dag} (directed acyclic graph) with a resolution refutation, where the nodes of the dag are the clauses in the proof; the sources (no incoming edges) of the dag are clauses in $F$ and each internal node has incoming edges from the at most two clauses used to deduce it.
A resolution refutation is \emph{tree-like} if every clause is used at most once as an input to the resolution rule; it is easy to see that the proof dag of a tree-like proof is a tree.

The main recurring theme that we will see in this survey is the use of \emph{total search problems} to understand problems in both proof and circuit complexity.
We now define a central family of total search problem that captures the complexity of proofs in many proof systems.
Given a CNF formula $F = C_1 \wedge \cdots \wedge C_m$ on $n$ variables, define the relation $\Search(F) \subseteq \B^n \times [m]$ by \[ (x, i) \in \Search(F) \Longleftrightarrow C_i(x) = 0\,.\]
We think of this relation as defining a search problem: given $x \in \B^n$ as input, find an $i \in [m]$ such that $(x, i) \in \Search(F)$.
It is easy to see that this search problem is \emph{total} (that is, $\forall x \exists i : (x, i) \in \Search(F)$) if and only if the CNF formula $F$ is unsatisfiable.
Moreover, it turns out that the complexity of the smallest \emph{decision tree} solving $\Search(F)$ captures exactly the complexity of tree-like resolution.
Indeed, it turns out that decision trees for $\Search(F)$ and tree-like resolution refutations of $F$ are essentially the same object!

\begin{theorem}
	\label{thm:treeres-dt}
	Let $F$ be an unsatisfiable CNF formula. There is a size-$s$, depth-$d$ tree-like resolution refutation of $F$ if and only if there is a size-$s$, depth-$d$ decision tree solving $\Search(F)$.
\end{theorem}

\subsection{Boolean formulas and Karchmer--Wigderson games}

Next we switch gears and recall a second famous result in complexity theory: the relationship between boolean circuit complexity and communication complexity due to Karchmer and Wigderson~\cite{Karchmer1990} (independently discovered by Yannakakis~\cite{Klawe1984}). We first define boolean circuits.

\begin{definition}
	A \emph{boolean circuit} on $n$ input variables is given by a sequence of functions $g_1, g_2, \ldots, g_s$ such that, for each $i \in [s]$, either 
	\begin{itemize}[noitemsep]
		\item $g_i = z_i$ or $g_i = \overline z_i$ for some input variable $z_i$, or
		\item $g_i = g_j \circ g_k$ where $\circ \in \set{\land, \lor}$ and $j, k < i$.
	\end{itemize}
	The boolean circuit has a natural underlying dag; we say the circuit is a \emph{formula} if the underlying dag is a tree.
	The size of the circuit is $s$, the number of gates, and we say that the circuit computes the boolean function $f$ if $g_s = f$.
	Finally, we say that the boolean circuit is \emph{monotone} if the negated input variables $\overline z_i$ are not used.
\end{definition}

In this language, one should think of boolean circuits as being analogous to resolution proofs, and boolean formulas as analogous to tree-like resolution proofs.
Both are, in some sense, ``deductive'' or ``bottom-up'': we start from ``atoms'' (either input clauses of the formula $F$, or variables of the underlying boolean function $f$), and deduce new intermediate objects from old ones using some predefined rules.

Extending this analogy, then, what corresponds to decision trees in the world of boolean circuit complexity?
Karchmer and Wigderson \cite{Karchmer1990} supplied an answer: \emph{communication protocols} solving \emph{Karchmer--Wigderson games}.

\begin{definition}
	Let $f: \B^n \rightarrow \set{0,1,*}$ be a (possibly partially defined) boolean function.
	The \emph{Karchmer--Wigderson game of $f$} is the relation $\KW(f) \subseteq f^{-1}(1) \times f^{-1}(0) \times [n]$
	defined by $(x, y, i) \in \KW(f)$ if and only if $x_i \neq y_i$.
	If $f$ is \emph{monotone} (meaning that for all $x, y \in \B^n$ if $x \leq y$ then $f(x) \leq f(y)$), then define the \emph{monotone Karchmer--Wigderson game of $f$}, denoted $\mKW(f) \subseteq f^{-1}(1) \times f^{-1}(0) \times [n]$, by $(x, y, i) \in \mKW(f)$ if and only if $x_i = 1, y_i = 0$.
\end{definition}

As in the case of $\Search(F)$ for unsatisfiable CNF formulas $F$, Karchmer--Wigderson games are also total search problems, when we view the pair $(x, y)$ as the input and $i \in [n]$ as the output.
Indeed, one should think of $\KW(f)$ as playing the same role as $\Search(F)$ does in proof complexity: it is a total search problem that in some sense captures the complexity of computing $f$.

Karchmer and Wigderson proved that this is true for \emph{boolean formulas}.
Indeed, in the same way that decision trees and tree-like resolution refutations are essentially identical objects, one can show that boolean formulas and communication protocols for $\KW(f)$ are also essentially identical.

\begin{theorem}[\cite{Karchmer1990}]\label{thm:kw-connection}
	Let $f : \B^n \rightarrow \set{0,1,*}$ be a partial boolean function.
	There is a depth-$d$ boolean formula computing $f$ if and only if there is a depth-$d$ deterministic communication protocol solving $\KW(f)$.
	Moreover, if $f$ is monotone, then there is a depth-$d$ monotone boolean formula computing $f$ if and only if there is a depth-$d$ communication protocol solving $\mKW(f)$.
\end{theorem}

The proof of the previous theorem is essentially identical to the proof of \autoref{thm:treeres-dt}, with objects in the world of resolution substituted for objects in the world of boolean circuits appropriately.
In \autoref{fig:dictionary}, we record a dictionary of the related objects in this correspondence.
This analogy is the lynchpin underlying the connections between proof and circuit complexity.
\begin{figure}[htbp]
	\centering
	\small
  	\begin{tabular}{ll|ll}
    \toprule
	  \bf Resolution proofs & \bf  Decision trees & \bf Boolean circuits & \bf Communication protocols \\ \midrule
	  unsatisfiable CNF $F$ & $\Search(F)$ & boolean functions $f$ & $\mKW(f)$ \\
	  clauses $C$ & subtrees & intermediate functions $g_i$ & rectangles $R$ \\
	  deduction rules & variable queries & circuit gates & communicated bits \\
    \bottomrule
	\end{tabular}
	\caption{The corresponding objects in resolution proofs and boolean circuits.}
	\label{fig:dictionary}
\end{figure}

\subsection{Feasible interpolation and lifting theorems}

Next we show how to use the above analogy to systematically relate the two worlds (proof and circuit complexity) together.
In order to do this, we will show how to directly relate the falsified clause search problem $\Search(F)$ to the monotone Karchmer--Wigderson game $\mKW(F)$.

Let us first answer a simple question.
Suppose $F = C_1 \wedge \cdots \wedge C_m$ is an unsatisfiable CNF formula on $n$ variables $Z = \set{z_1, z_2, \ldots, z_n}$.
How can we turn the search problem $\Search(F) \subseteq \B^n \times [m]$ into a two-party communication task?
The obvious way is to take a partition of the variables $Z = X \cup Y$, and then give the assignment to the $X$-variables to Alice and the assignment to the $Y$-variables to Bob.
Let us call this two-party search problem $\Search^{(X, Y)}(F)$, and observe that an efficient decision tree for $\Search(F)$ immediately implies an efficient communication protocol for $\Search^{(X, Y)}(F)$ simply by simulating the decision tree.

This is progress, but while $\Search^{(X, Y)}(F)$ is a total search problem it is not a Karchmer--Wigderson game.
It turns out, however, that we can interpret it as the \emph{monotone} Karchmer--Wigderson game for some partial boolean function.
The idea for this goes back to Razborov \cite{Razborov1990}, although it has been reintroduced in the recent works on lifting theorems \cite{GoosP2018, GPW2018, FPPR2017}, as well as independently by Hrube\v{s} and Pudl\'{a}k \cite{HP2017} as a generalisation of the classic theory of \emph{feasible interpolation} introduced by Kraj\'{i}\v{c}ek~\cite{Krajicek1997}.
We will describe the definition introduced by Hrube\v{s} and Pudl\'{a}k.
If $C$ is a clause and $X$ is a subset of its variables we let $C^X$ denote the subclause of $C$ containing only literals over $X$.
\begin{definition}
	\label{def:unsat-certificate}
	Let $F = C_1 \wedge \cdots \wedge C_m$ be an unsatisfiable CNF formula and let $(X,Y)$ be any partition of its variables. The \emph{unsatisfiability certificate} associated with $F$ and $(X,Y)$ is the partial monotone function $\cert_{F}^{(X,Y)} :  \{0,1\}^m \rightarrow \{0,1,*\}$ defined as 
	\[ 
	\cert_{F}^{(X,Y)}(z) := \begin{cases} 
		1 &\mbox{if $\displaystyle \bigwedge_{i: z_i = 0} C_i^X$ is satisfiable}, 	\\
		0 &\mbox{if $\displaystyle \bigwedge_{i: z_i = 1} C_i^Y$ is satisfiable}, \\
		* & \text{otherwise.}
 	\end{cases}
 	\]
	We may suppress the partition $(X, Y)$ or the underlying CNF formula $F$ in $\cert_F^{(X,Y)}$ when it is clear from context.
\end{definition}

Indeed, one can show that $\mKW(\cert^{(X, Y)}_F)$ is communication reducible to $\Search^{(X, Y)}(F)$ and \emph{vice-versa}.
Using the unsatisfiability certificate, we can therefore prove the following ``feasible interpolation'' theorem relating the boolean formula depth of the unsatisfiability certificate to the resolution depth of the underlying CNF formula.

\begin{theorem}
	Let $F$ be an unsatisfiable CNF formula, let $(X, Y)$ be any partition of the variables of $F$, and let $\cert_F$ denote the unsatisfiability certificate with respect to this variable partition.
	If there is a resolution proof of depth $d$ refuting $F$ then there is a monotone boolean formula of depth~$d$ computing $\cert_F$.
\end{theorem}

First note that the previous theorem implies that \emph{upper bounds} on proofs imply \emph{upper bounds} on circuits; thus, conversely, lower bounds on monotone circuit complexity implies lower bounds on proof complexity.
Second, although we have stated the previous theorem in terms of depth, the argument can be easily modified to capture the \emph{size} of the two systems as well.

Finally, it is natural to ask if a converse to the previous result can hold.
That is, are there unsatisfiable CNF formulas $F$ such that the decision tree complexity of solving $\Search(F)$ is a lower bound on the communication complexity of solving $\mKW(\cert_F)$?
The answer to this is an emphatic \emph{yes}, thanks to well-known \emph{query-to-communication lifting theorems}, which have been proved in a long line of work starting with Raz and McKenzie~\cite{Raz1999}.
These powerful and general results allow us to generate entire families of formulas $F$ for which the decision tree complexity of $\Search(F)$ and the communication complexity of $\mKW(\cert_F)$ are tightly related.

The basic idea of a lifting theorem is simple: we start with an unsatisfiable CNF formula $F$, and replace each variable $z_i$ in $F$ with a \emph{gadget function} $g(x_i, y_i)$ on new ``fresh'' inputs $x_i, y_i$.
In this way, we obtain a new formula $F \circ g^n = F(g(x, y))$, which we can transform back into a CNF formula using syntactic manipulations.
Perhaps the most often used example of a gadget function $g$ is the \emph{index gadget} $\IND_\ell : [\ell] \times \B^{\ell} \rightarrow \B$, which takes in a string $y \in \B^\ell$ and a pointer $x \in [\ell]$ to a bit in the string and outputs $\IND_{\ell}(x, y) = y_x$.
Raz and McKenzie~\cite{Raz1999} (further developed by~\cite{GPW2018,CKLM2019}) proved the following lifting theorem that relates the query complexity of $\Search(F)$ to the communication complexity of $\mKW(\cert_{F \circ \IND})$.

\begin{theorem}\label{thm:det-lifting}
	Let $F$ be an unsatisfiable CNF formula on $n$ variables, and let $m = n^C$ for a large enough constant $C \geq 1$.
	The size of any monotone boolean formula computing $\cert_{F \circ \IND_m}$ is at least $n^{\Omega(D(F))}$, where $D(F)$ is the minimum depth of any (tree-like) resolution refutation of $F$.
\end{theorem}

This result allows us to deduce lower bounds on monotone formula complexity (a difficult task) automatically from resolution depth lower bounds (a relatively simple task).
Moreover, by choosing the underlying unsatisfiable CNF formulas $F$ to be specially structured, we can obtain lower bounds for many standard functions considered in circuit complexity, such as the \emph{$st$-connectivity}, \emph{clique} and \emph{generation} functions.

The $st$-connectivity function $\stconn_n$ takes as input 
an $n$-vertex directed graph with two
distinguished vertices~$s$ and $t$, and outputs $1$ if there is a directed path from $s$ to~$t$ and $0$ otherwise.
Karchmer and Wigderson~\cite{Karchmer1990} proved, for the case of undirected graphs,
that any monotone boolean formula computing $\stconn_n$ must have size $n^{\Theta(\log n)}$, and alternative proofs were given by \cite{Grigni1995,Potechin2017,Robere2018} for the case of directed graphs.
A similar lower bound follows from \autoref{thm:det-lifting} by considering $F$ to be the induction principle of~\cite{Buss1998}
(see \cite{Robere2018} for a proof).

The clique function $\clique_{n,k}$ takes as input an $n$-vertex graph and outputs $1$ if it contains a $k$-clique and $0$ otherwise. It is not hard to see that
there is a monotone boolean formula of size $n^{O(k)}$ that computes $\clique_{n,k}$ by checking all $\binom{n}{k}$~potential $k$-cliques.
Razborov~\cite{Razborov1985} proved the first superpolynomial size lower bound for monotone circuits precisely for the clique function. This was later improved in~\cite{AlonBoppana1987,WegenerBook1987} to an $n^{\Omega(\sqrt{k})}$ lower bound for $k \leq n^{2/3 - \epsilon}$ and more recently to an $n^{\Omega({k})}$ lower bound for $k \leq n^{1/3 - \epsilon}$~\cite{Cavalar2020}. 
For monotone formulas, a weaker lower bound but with a simpler proof was presented in~\cite{Goldmann1992} and this was improved to an $\exp(\Omega(k))$ lower bound for $k\le\frac{2}{3}n+1$ in~\cite{Raz1992}. Raz and McKenzie~\cite{Raz1999} obtained an $n^{\Omega(k)}$ size lower bound for monotone formulas for $k \leq n^{\epsilon}$ by applying \autoref{thm:det-lifting} with $F$ being a certain formulation of the pigeonhole principle. 

The generation function~$\gen_n$ was introduced by Raz and McKenzie~\cite{Raz1999}
in order to separate the monotone $\NC$ hierarchy.
It is defined as follows.
Given a set $\mathcal{T}\subseteq\left[n\right]^{3}$, we say
that $\mathcal{T}$ \emph{generates} a point $w\in\left[n\right]$ if $w=1$,
or if there is a triplet $(u,v,w)\in\mathcal{T}$ such that $\mathcal{T}$ generates
$u$ and~$v$. The function~$\gen$ takes as an input
a set~$\mathcal{T}\subseteq\left[n\right]^{3}$ and outputs $1$ if $\mathcal{T}$ generates~$n$ and $0$ otherwise.
By applying \autoref{thm:det-lifting} to the so-called \emph{pebbling formulas}, Raz and McKenzie~\cite{Raz1999} proved an $\exp(\Omega(n^\epsilon))$ monotone formula size lower bound for $\gen_n$. Since $\gen_n$ is computable by a polynomial-size monotone circuit, this implies that $\mNC \neq \mP$. 
By a similar argument, they also showed that $\mNC^{i} \neq \mNC^{i+1}$. 
A later extension of this lifting theorem in~\cite{dRNV2016}, building on~\cite{GPW2018}, completed the picture
of the relation between the $\mAC$ and $\mNC$ hierarchies, which
looks like $\mNC^{i}  \subsetneq \mAC^{i} \subsetneq \mNC^{i+1}$.

We summarise the main formula size lower bounds that follow from \autoref{thm:det-lifting} as follows.

\begin{corollary}
Let $\epsilon$ be a small enough constant. 
The size of any monotone boolean formula computing:
\begin{enumerate}[noitemsep]
\item $\stconn_n$ is at least $n^{\Omega(\log n)}$;
\item $\clique_{n,k}$  for $k\leq n^\epsilon$ is at least $n^{\Omega(k)}$; and
\item $\gen_n$ is at least $\exp(n^\epsilon)$.
\end{enumerate} 
\end{corollary}

Recent extensions of \autoref{thm:det-lifting}~\cite{CKLM2019,LMMPZ20} in particular improve the size of the gadget and as a consequence the parameter $\epsilon$ above need not be very small. We also note that the best known monotone formula lower bounds for these functions follow from another lifting theorem~\cite{PR2017,dRMNPRV2020}.

Let us conclude with three main takeaway lessons from this section.

\begin{itemize}
	\item First, the structure of tree-like resolution proofs and boolean formulas are extremely similar, and this shared structure can be exploited in a nice way using the appropriate total search problems.
	\item Second, by exploiting this analogy, we can easily deduce generic \emph{feasible interpolation} results, which allow us to translate circuit lower bounds into proof lower bounds.
	\item Third, by using \emph{composition}, one can prove a lifting theorem that is essentially a \emph{converse} to feasible interpolation. 
		In concrete terms this means that, for certain tasks, lower bounds on proofs imply lower bounds on boolean circuits.
\end{itemize}

In the remainder of this survey we will see several variations on this theme that have occurred in the recent literature.
Next, we will examine how to move from \emph{tree-like} proofs and circuits to more general \emph{dag-like} models.

\section{Dag-Like Resolution Proofs and Monotone Boolean Circuits} \label{sec:dags}

In order to capture dag-like models, we will need to introduce dag-like versions of decision trees and communication protocols that capture resolution and boolean circuits, respectively.
Both of our dag-like models will share the following definition.\footnote{For those who are familiar, the next definition is closely related to the notion of a ``semantic derivation'', introduced by Kraj\'{i}cek \cite{Krajicek1997}.}

\begin{definition}
	Let $\mathcal{S} \subseteq \mathcal{I} \times \mathcal{O}$ be a total search problem with input $\mathcal{I}$ and output $\mathcal{O}$, and let $\mathcal{F}$ be a family of functions from $\mathcal{I} \rightarrow \B$.
	An \emph{$\mathcal{F}$-dag solving $\mathcal{S}$} is given by a directed acyclic graph of fan-out at most $2$ such that each node $v$ in the graph is associated with a function $f_v \in \mathcal{F}$ satisfying the following:
	\begin{itemize}
		\item \emph{Root.} There is a unique root node $r$ with fan-in $0$ associated with the (trivial) function $1$.
		\item \emph{Internal Consistency.} For each internal node $v$ with children $u, w$ we have that \[f_v^{-1}(1) \subseteq f_u^{-1}(1) \cup f_w^{-1}(1)\,.\]
		\item \emph{Solutions.} Each leaf node $v$ is labelled with an $o \in \mathcal{O}$ such that for all $x \in f_v^{-1}(1)$, $(x, o) \in \mathcal{S}$.
	\end{itemize}
	The \emph{size} of an $\mathcal{F}$-dag is the number of nodes in the graph. 
\end{definition}

First we instantiate this model and consider a dag-like version of decision trees, which we call \emph{conjunction-dags} following \cite{GGKS2020}.
This model and close variations on it have been studied in the literature in a number of separate works prior to \cite{GGKS2020}; notably, by Pudl\'{a}k \cite{Pudlak2000}, and then by Atserias and Dalmau \cite{AtseriasD2008}, under the name \emph{Prover-Adversary games}.
\begin{definition}
	Let $F$ be an unsatisfiable CNF formula.
	A \emph{conjunction-dag} solving $\Search(F)$ is an $\mathcal{F}$-dag solving $\Search(F)$ where $\mathcal{F}$ is the family of all conjunctions over boolean literals in $F$.
\end{definition}

A few remarks on this definition are in order.
First, it is not hard to interpret the model as a dag-like version of a decision tree (indeed, a decision tree is a tree-like conjunction-dag!).
Given an input $x \in \B^n$, the ``search algorithm'' starts by examining the root node, at which $x$ trivially satisfies the associated conjunction.
By the internal consistency property, if $x$ satisfies the conjunction at a node $v$, then it also satisfies the conjunction at one of the children of $v$, and thus the algorithm can proceed from $v$ to the satisfied child.
Finally, once the algorithm reaches a leaf node, the final property guarantees that the leaf will be labelled with a solution to $\Search(F)$.

Second, it is also easy to see how this model corresponds to resolution.
If we are given a resolution proof $\Pi$, then we can simply replace each clause in $\Pi$ with the negation of that clause (i.e.,~a conjunction); it is easy to verify that the resulting dag satisfies the three properties of a conjunction-dag.
The converse direction is not much harder; we refer to \cite{GGKS2020} for a proof. 

\begin{theorem}
	\label{thm:res-conjdag}
	Let $F$ be an unsatisfiable CNF formula. There is a size-$s$ resolution refutation of $F$ if and only if there is a size-$s$ conjunction-dag solving $\Search(F)$.
\end{theorem}

Now, we can similarly define a dag-like model of two-party communication that corresponds to boolean circuits.
The model of dag-like communication was first introduced by Razborov \cite{Razborov1995}, where it was shown that dag-like communication captures boolean circuit size in the same way that deterministic communication captures boolean formula size.
Razborov's original definition used the language of $\TFNP$, particularly as a ``communication version'' of the complexity class $\PLS$ (see \autoref{sec:perspectives} for more details on this).
The model was subsequently studied by Kraj\'i\v{c}ek~\cite{Krajicek1997} and simplified by Sokolov \cite{Sokolov2017}, although we will use the following version from \cite{GGKS2020}.
As we will be primarily interested in monotone circuit complexity, we define it purely for the $\mKW$ game.

\begin{definition}
	Let $f: \B^n \rightarrow \set{0,1,*}$ be a (possibly partial) monotone boolean function.
	A \emph{rectangle-dag} solving $\mKW(f) \subseteq f^{-1}(1) \times f^{-1}(0) \times [n]$ is an $\mathcal{F}$-dag solving $\mKW(f)$ where $\mathcal{F}$ is the family of all indicator functions of combinatorial rectangles over $f^{-1}(1) \times f^{-1}(0)$.
\end{definition}

It is again easy to observe that standard deterministic communication protocols for $\mKW(f)$ are rectangle-dags where the underlying graph is a tree, and furthermore that one can associate these objects with a top-down ``search algorithm'' as we described for conjunction-dags above.

Second, as we stated above, it is known that the size of rectangle-dags captures boolean circuit size.
While we will state it for monotone circuits, a similar result also holds for non-monotone circuits.

\begin{theorem}[\cite{Razborov1995, Sokolov2017}]
	Let $f$ be a monotone boolean function.
	There is a size-$s$ monotone boolean circuit computing $f$ if and only if there is a size-$s$ rectangle-dag solving $\mKW(f)$.
\end{theorem}

Finally, just as in the previous section, one can use these models to prove a simple ``feasible interpolation''-style result for resolution proofs in terms of monotone boolean circuits.
This feasible interpolation result was originally observed for certain structured formulas by Kraj\'i\v{c}ek~\cite{Krajicek1997} and Pudl\'ak~\cite{Pudlak1997} using direct arguments; one of the benefits of our abstract setup is that we get a similar feasible interpolation result in terms of the unsatisfiability certificate immediately and transparently from the definitions.

\begin{theorem}
	Let $F = C_1 \wedge \cdots \wedge C_m$ be an unsatisfiable CNF formula, let $(X, Y)$ be any partition of the variables of $F$, and let $\cert_F$ denote the corresponding unsatisfiability certificate.
	If there is a resolution refutation of $F$ of size $s$ then there is a monotone boolean circuit computing $\cert_F$ of size $O(s)$.
\end{theorem}

As in the case of tree-like proofs and boolean formulas, it is natural to consider whether or not a \emph{converse} to the previous result is possible.
If so, this would allow us to translate lower bounds for resolution proofs (which is a very well-developed theory, e.g.,~\cite{Haken1985, BenSassonW2001}) into lower bounds for monotone boolean circuits, for which we essentially have only one technique, the \emph{method of approximations} with its variations \cite{Razborov1985, Jukna97, HarnikR00, BergU1999}.
Indeed, such a result was proven by G\"{o}\"{o}s, Garg, Kamath and Sokolov \cite{GGKS2020}, who crucially used the search viewpoint described above in their proof.

\begin{theorem}[\cite{GGKS2020}]\label{thm:ggks-lifting}
	Let $F$ be an unsatisfiable CNF formula on $n$ variables, and let $m = n^C$ for a sufficiently large constant $C$.
	Then the size of any monotone boolean circuit computing $\cert_{F \circ \IND_m}$ is at least $n^{\Omega(w(F))}$, where $w(F)$ is the minimum width of any resolution refutation of~$F$.
\end{theorem}

Note that the measure being ``lifted'' here is resolution refutation \emph{width}, in contrast to \autoref{thm:det-lifting} where it was \emph{depth}. This difference is indeed necessary, since  both \stconn\ and \gen\ have polynomial size monotone circuit. Given that there is a width lower bound for refuting the pigeonhole principle, by applying \autoref{thm:ggks-lifting} in the same way as done in~\cite{Raz1999}, we get a monotone circuit lower bound for \clique. This lower bound, as is the case for most other clique lower bounds, actually holds for (any extension of) the partial function \emph{clique-colouring}. The function $\cliquecol_{n,k}$ receives as input an $n$-vertex graph and outputs $1$ if it has a $k$-clique, $0$ if it is $(k-1)$-colourable and can output anything if neither is the case. Tardos~\cite{Tardos1988} proved that there is an extension of \cliquecol\ that is monotone and is in \P, and therefore this gives an exponential separation between monotone and non-monotone circuits.

Another interesting application of \autoref{thm:ggks-lifting}, observed in a follow-up work~\cite{GKRS2019}, is that it improves this separation to an exponential separation between $\NC^2$ and \mP. This is done by considering the monotone function $3\xorsat_n$, which receives as input an indicator vector for a set of $3\xor$ constraints over $n$ variables and outputs $1$ if the set is unsatisfiable and $0$ otherwise. By lifting so-called \emph{Tseitin formulas}, which are known to require large resolution width to be refuted~\cite{Urquhart1987}, they obtain a monotone circuit lower bound for $3\xorsat$. 
Since $3\xorsat$ is in $\NC^2$~\cite{Mulmuley1987}, this improves on the exponential monotone vs.\ non-monotone separation.

\begin{corollary}
Let $\epsilon$ be a small enough constant. The size of any monotone boolean circuit computing:
\begin{enumerate}[noitemsep]
\item $\clique_{n,k}$ for $k\leq n^\epsilon$ is at least $n^{\Omega(k)}$; and
\item $3\xorsat_n$ is at least $\exp(n^{\epsilon})$.
\end{enumerate} 
\end{corollary}

As we will discuss in the next section, \autoref{thm:ggks-lifting} can also be extended to obtain lower bounds for \emph{monotone real circuits}, and therefore \emph{cutting planes proofs}.
With regards to monotone circuit lower bounds, however, it is natural to wonder how the lower bounds that follow from the previous theorem compare to the known results provable via the method of approximations.
It seems that the techniques are ultimately incomparable, but recent work of Lovett et~al.~\cite{LMMPZ20} gave an alternative proof of \autoref{thm:ggks-lifting} using the \emph{sunflower lemma}, which is the same central combinatorial tool used in the method of approximations.

\section{Cutting Planes and Real Monotone Circuits}
\label{sec:cutting-planes}

Cutting planes is a proof system that captures integer programming reasoning, based on the \emph{cutting planes method} of Gomory \cite{Gomory1963} and Chv\'{a}tal \cite{Chvatal73a}.
In the same way as the basic unit in resolution are clauses,
in cutting planes they are integer linear inequalities.
Let~$\mathcal{I}$ be a system of unsatisfiable integer linear inequalities over variables $x_1,\ldots,x_n$.
A \emph{cutting planes refutation} of~$\mathcal{I}$
is a sequence of linear inequalities $(L_1, \ldots, L_{\ell})$ such that 
$L_{\ell}$ is $0\geq 1$ and each $L_i$ is either in~$\mathcal{I}$
or can be inferred from previous inequalities by one of the two following deduction rules:
\begin{align*}
\text{linear combination}&
\enspace
\AxiomC{$\sum_i a_i x_i \geq A $}
\AxiomC{$ \sum_i b_i x_i \geq B$}
\BinaryInfC{$\sum_i \, (\alpha a_i +  \beta b_i) x_i \geq \alpha A + \beta B$}
\DisplayProof\mbox{~, or } 
\text{division}
\enspace
\AxiomC{$\sum_i \alpha a_i x_i \geq A$}
\UnaryInfC{$\sum_i a_i x_i \geq \lceil{A/\alpha\rceil}$}
\DisplayProof~,
\end{align*}
where $a_i,b_i,A,B,\alpha,\beta$ are integers and $\alpha,\beta$ are positive.
The \emph{length} of the refutation is $\ell$. 
We say that the refutation has \emph{bounded coefficients} if all
coefficients are bounded in absolute value by some polynomial in $n$.
As before, we say the 
refutation is \emph{tree-like} if the 
underlying proof dag is a tree.

For most of what we will discuss here, the exact rules of derivation are not particularly
important, as long as they are sound over integers and refutationally complete. 
It is convenient, therefore, to define a \emph{semantic cutting planes refutation} of $\mathcal{I}$ as 
a sequence of linear inequalities $(L_1, \ldots, L_{\ell})$ such that 
$L_{\ell}$ is $0\geq 1$ and each $L_i$ is either in~$\mathcal{I}$
or is \emph{semantically implied} (over boolean variables) by two previous inequalities $L_j$ and $L_k$ where $j,k < i$, i.e., all boolean points that satisfy both $L_j$ and~$L_k$ also satisfy $L_i$.

In order to use cutting planes to refute unsatisfiable CNF formulas, we 
must consider the translation of clauses into linear inequalities.
A clause $\bigvee_{i\in P} x_i \lor \bigvee_{i\in N} \overline{x}_i$
is translated to $\sum_{i\in P} x_i + \sum_{i\in N} (1 - x_i) \geq 1$,
where we interpret $0$ as false and $1$ as true,
and a CNF formula $F$ over variables $x_1,\ldots,x_n$
is translated into a system of linear inequalities consisting
of the translation of each clause in $F$ and the inequalities $x_i \geq 0$
and $x_i \leq 1$ for $i\in\{1,\ldots,n\}$ to guarantee variables take boolean
values. We often identify $F$ with this system of linear inequalities.

As was the case in resolution, we can view a semantic cutting planes refutation
in a top-down manner, as an analogue of decision trees. 
The difference is that, in order to capture cutting-planes, we need dags where nodes are associated to halfspaces.

\begin{definition}
	Let $F$ be an unsatisfiable CNF formula.
	A \emph{halfspace-dag} solving $\Search(F)$ is an $\mathcal{F}$-dag solving $\Search(F)$ where $\mathcal{F}$ is the family of all halfspaces over the boolean variables in $F$.
\end{definition}

Following precisely the same argument that was sketched in \autoref{sec:dags} for resolution and conjunction-dags, it is possible to prove an equivalence between cutting planes refutations of $F$ and halfspace-dags solving $\Search(F)$.

\begin{theorem}
	\label{thm:cp-dt}
	Let $F$ be an unsatisfiable CNF formula. There is a size-$s$ semantic cutting planes refutation of $F$ if and only if there is a size-$s$ halfspace-dag solving $\Search(F)$.
\end{theorem}

\subsection{Monotone real circuits and triangle-dags} 

In order to extend Kraj\'i\v{c}ek's interpolation technique~\cite{Krajicek1997}
to cutting planes, Pudl\'ak~\cite{Pudlak1997} defined
\emph{monotone real circuits} as a generalisation of monotone boolean circuits where each gate is allowed to
compute any non-decreasing real function of its inputs, but the inputs and the output of the circuit are boolean.

Soon after, Kraj\'i\v{c}ek~\cite{Krajicek1998} defined \emph{monotone protocols}
with the goal of extending the interpolation technique to a wider class of proof systems. 
Here we define a subclass of these monotone protocols 
that we will refer to as \emph{triangle-dags}\footnote{In~\cite{Krajicek1998} this subclass is called \emph{monotone protocols whose real communication complexity is $1$}; in~\cite{Sokolov2017} \emph{real communication games} and in~\cite{HP2018} \emph{real protocols}.} as coined in~\cite{GGKS2020}.

We say $T\subseteq \mathcal{X} \times \mathcal{Y}$ is a \emph{combinatorial triangle} if for all $(x,y) \in T$ and $(x',y') \in T$ it holds that either $(x,y')\in T$ or $(x',y) \in T$ (or both). In other words, there is some labelling
of the rows $a_T : \mathcal{X} \rightarrow \mathbb{R}$ and
of the columns $b_T : \mathcal{Y} \rightarrow \mathbb{R}$ such that
$T = \{(x,y) : a_T(x) < b_T(y)\}$. In particular, note that every combinatorial rectangle is a triangle.

\begin{definition}
	Let $f: \B^n \rightarrow \set{0,1,*}$ be a (possibly partial) monotone boolean function.
	A \emph{triangle-dag} solving $\mKW(f) \subseteq f^{-1}(1) \times f^{-1}(0) \times [n]$ is an $\mathcal{F}$-dag solving $\mKW(f)$ where $\mathcal{F}$ is the family of all indicator functions of combinatorial triangles over $f^{-1}(1) \times f^{-1}(0)$.
\end{definition}

We once again have a tight relation between a circuit model and an $\mathcal{F}$-dag.

\begin{theorem}[\cite{HP2018}]\label{thm:mKW-real}
Let $f$ be a monotone boolean function.
There is a monotone real circuit of size $s$ computing $f$ if and only if
there is a triangle-dag of size $s$ solving $\mKW(f)$.
\end{theorem}

Indeed, just as we thought of rectangles as the communication ``analogue'' of clauses, we think of triangles as the communication ``analogue'' of linear inequalities.
The fact that small monotone real circuits for $f$ give small triangle-dags
for the \mKW-game on~$f$ was already observed by Kraj\'i\v{c}ek~\cite{Krajicek1998}.
The other direction is less obvious and was shown more recently by
Hrube{\v{s}} and Pudl{\'{a}}k~\cite{HP2018}.

Some of the known lower bounds for monotone boolean circuits
were generalised 
to monotone real circuits: clique-colouring~\cite{Pudlak1997}, broken mosquito screen~\cite{Haken1999}, perfect matching~\cite{Fu1997}, $st$-connectivity~\cite{Johannsen1998}, generation~\cite{BEGJ00}.
Jukna~\cite{Jukna1999} gave a general lower bound criterion for monotone boolean circuits and then showed that the same criterion also holds for monotone real circuits. 

These results might suggest that monotone real circuits cannot be much more powerful than monotone boolean circuits. Rosenbloom~\cite{Rosenbloom1997}, however, proved the contrary. He showed that monotone real circuits 
can be exponentially stronger than even \emph{non-monotone} boolean circuits. The proof is simple: by a counting argument, most slice functions cannot be computed by boolean circuits of subexponential size, whereas any slice function can be computed by a linear-size, logarithmic-depth monotone real formula.
This implies, together with the clique-colouring lower bound~\cite{Pudlak1997} and the fact that there is a monotone function in $\Ppoly$ that extends clique-colouring~\cite{Tardos1988}, that monotone real circuits and non-monotone boolean circuits are incomparable.

\subsection{Feasible interpolation and lifting theorems}

Kraj\'i\v{c}ek's interpolation technique~\cite{Krajicek1997} 
gives a method of obtaining resolution lower bounds from monotone circuit lower bounds.
The technique is actually more general and implies lower bounds
also for cutting planes with bounded coefficients. 
Pudl\'ak~\cite{Pudlak1997} was able to extend these results to unrestricted cutting planes by showing that cutting planes refutations can be interpolated by monotone real circuits.
The classical method of interpolation only applies to certain structured formulas but, as we have discussed in \autoref{sec:tree-like}, they can be generalised to other formulas by considering unsatisfiability certificates \cite{HP2017, FPPR2017}.

\begin{theorem}\label{thm:cp-interpolation-HP}
	Let $F$ be an unsatisfiable CNF formula with $m$ clause, let $(X, Y)$ be any partition of the variables of $F$, and let $\cert_F$ denote the corresponding unsatisfiability certificate.
	If there is a cutting planes refutation of $F$ of length $s$ then there is a monotone real circuit computing $\cert_F$ of size $O(s + m^2)$.
\end{theorem}

Pudl\'ak's interpolation theorem~\cite{Pudlak1997} implied the first superpolynomial lower bound for unrestricted cutting planes and for two decades monotone interpolation was the only known method for proving cutting planes lower bounds.
Some of the formulas for which we can obtain cutting planes lower bound via this monotone interpolation theorem include
clique-colouring~\cite{Pudlak1997}, 
random $k$-CNF formulas for $k = \Theta(\log n)$~\cite{HP2017,FPPR2017}, and weak bit pigeonhole principle~\cite{HP2017}.

Again we can consider if a converse of this theorem holds. 
Can we obtain monotone real circuit lower bounds from proof complexity lower bounds?
The answer is once again \emph{yes}. The dag-like lifting theorem by G\"o\"os et al.~\cite{GGKS2020} holds also for triangle-dags and therefore for monotone real circuits. Note that the proof complexity measure that is being lifted is still resolution width.

\begin{theorem}[\cite{GGKS2020}]
	Let $F$ be an unsatisfiable CNF formula on $n$ variables, and let $m = n^C$ for a sufficiently large constant $C$.
	Then the size of any monotone real circuit computing $\cert_{F \circ \IND_m}$ is at least $n^{\Omega(w(F))}$, where $w(F)$ is the minimum width of any resolution refutation of~$F$.
\end{theorem}

This implies that the lower bounds for monotone boolean circuits that can be obtained via this lifting theorem 
also hold for monotone real circuits.
The main motivation for proving this lifting theorem was not, however, to obtain monotone real circuit lower bounds, but to obtain cutting planes lower bounds. 
The lifting theorem, as stated in~\cite{GGKS2020}, gives a lower bound for triangle-dags. It is straightforward to see that every halfspace-dag is, in particular, a triangle-dag. 
Therefore, in order to obtain a cutting planes lower bound it is sufficient to reduce the $\mKW$ game on $\cert_{F \circ \IND_m}$ to $\Search(F')$ for some formula $F'$. This can be done via syntactic manipulations, giving rise to the so-called \emph{lifted formulas}, or with some more clever reductions, which can give lower bounds for more natural formulas such as the clique-colouring formula. 
Lifting theorems significantly increase the families of formulas for which we have cutting planes lower bounds: any formula for which we can prove a strong-enough resolution width lower bound can be transformed into a formula that is hard for cutting planes.

\begin{corollary}[\cite{GGKS2020}]
Given any unsatisfiable $k$-CNF formula $F$ on $n$ variables, there is a polynomial-time constructible unsatisfiable $2k$-CNF formula $F'$ on $n^{O(1)}$ variables such that any cutting planes refutation of $F'$ must have length at least $n^{\Omega(w(F))}$, where $w(F)$ is the minimum width of any resolution refutation of~$F$.
\end{corollary}

Prior to~\cite{GGKS2020}, 
lifting theorems were proven for a related model called
\emph{\rcprotocols}~\cite{Krajicek1998,BEGJ00}.
In this setting, in order to solve a relation or search problem $S\subseteq \mathcal{X} \times \mathcal{Y} \times \mathcal{O}$,
Alice and Bob communicate via a referee that acts as a greater-than oracle: at each round,
Alice and Bob each send a real number to the referee and the referee replies to both player
with a bit indicating whether Alice's number is greater than Bob's number. 

Such a protocol can be viewed as a binary tree where every non-leaf node $v$ is labelled
by two functions $f^A_v : \mathcal{X} \rightarrow \mathbb{R}$ and
$f^B_v : \mathcal{Y} \rightarrow \mathbb{R}$, and
every leaf node $u$ is labelled by some output $o_u \in \mathcal{O}$. 
For an input $(x, y) \in  \mathcal{X} \times \mathcal{Y}$, 
the output of the protocol is defined by constructing a path from the root to a leaf according to the following rule: at node $v$ if $f^A_v(x) > f^B_v(y)$ then the path continues to the left child of $v$, otherwise the path continues to the right child of $v$. When the path reaches a leaf $u$ of the tree, then the output is $o_u$. We say the protocol computes $S$ if for every $(x, y) \in  \mathcal{X} \times \mathcal{Y}$ the output $o$ of the protocol is such that $(x,y,o) \in S$.
The depth of the protocol is the length of the longest root-to-leaf path and the size is the number of nodes in the tree.

It is not hard to see that this model can be quite powerful:
the equality function---which can only be solved by exponential-size
triangle-dags---can be trivially solved by constant-size \rcprotocols. Nevertheless, Bonet et al.~\cite{BEGJ00} showed that it is possible to prove lower bounds for this model by extending~\cite{Raz1999}. 
Note that this in particular implies
a tree-like version of the lifting theorem in~\cite{GGKS2020}.

\begin{theorem}[\cite{BEGJ00}]
	Let $F$ be an unsatisfiable CNF formula on $n$ variables, and let $m \geq n^C$ for a sufficiently large constant $C$.
	Any \rcprotocol that solves $\Search(F)\circ \IND_m$ must have depth at least ${\Omega(D(F)\cdot\log m})$, where $D(F)$ is the minimum depth of any resolution refutation of~$F$.
\end{theorem}

We note that $\Search(F)\circ \IND_m$ is communication reducible to the \mKW\ game for $\cert_{F \circ \IND_m}$.
Thus, following Raz and McKenzie~\cite{Raz1999}, this lifting theorem implies that the \mKW\ game for the generation function---which can be solved by polynomial-size rectangle-dags---requires exponential-size \rcprotocols. 
From this result, Bonet et al.~\cite{BEGJ00}
use the fact~\cite{Krajicek1998,Johannsen1998} that a lower bound for \rcprotocols 
of the \mKW\ game of a function $f$
implies a lower bound for monotone real formulas of $f$
to show that monotone boolean circuits
can be exponentially stronger than monotone real formulas (the opposite separation follows from~\cite{Rosenbloom1997}). 

Bonet et al.~\cite{BEGJ00} then use this result to 
show that resolution can be exponentially stronger than tree-like
cutting planes (the opposite separation is witnessed by the pigeonhole
principle). Their proof uses their circuit separation and Pudl\'ak's interpolation theorem~\cite{Pudlak1997}, but 
it is possible to obtain the cutting planes lower bound directly from the lower bound for \rcprotocols.

In the wake of G\"o\"os et al.~\cite{GPW2018}, 
de Rezende et al.~\cite{dRNV2016} extended this lifting theorem to a round-preserving setting. 
This was used to obtain length-space trade-offs for cutting planes. We note that, in contrast to~\cite{BEGJ00}, for this application it is indeed important that the communication model of the lifting theorem is \rcprotocols and not triangle-dags.

\section{Perspectives and Open Problems}
\label{sec:perspectives}

In this final section, we discuss directions for further research inspired by the themes of this survey. Our main conceptual message is this: We advocate to foster the \emph{interplay} between the different subfields of complexity theory---circuit complexity, proof complexity, query/communication complexity, and the theory of total search problems. Firstly, can we find more connections between these theories? To discover such interconnections, we adhere to the following philosophy.
\begin{center}\itshape
\begin{tabular}{rcl}
\bf\itshape Guiding philosophy: && \itshape A natural concept introduced in one theory should have\\
&&\itshape  a natural counterpart in another theory.
\end{tabular}
\end{center}
Secondly, given such connections, can we bring the techniques of one field to bear on the open questions of another field?

\subsection{Query/communication \TFNP}

In the context of classical (Turing machine) complexity theory, the study of total $\NP$ search problems ($\TFNP$) was initiated by Megiddo and Papadimitriou~\cite{Megiddo1991,Papadimitriou1994}. This theory aims to characterise the complexities of search problems that have a solution for every input and where a given solution can be efficiently checked for validity. By now, this theory has flowered into a sprawling jungle of widely-studied complexity classes (such as $\PLS$~\cite{Johnson1988}, $\PPA/\PPAD/\PPP$~\cite{Papadimitriou1994}, $\CLS$~\cite{Daskalakis2011}) that serve to classify the complexities of many important search problems. For instance, one of the most celebrated results in the field is that computing a Nash equilibrium (every game must have one) is complete for $\PPAD$~\cite{Daskalakis2009,Chen2009}.

\begin{example}
As a running example, we recall the class $\PLS$ (Polynomial Local Search) which embodies the combinatorial principle \emph{``every dag has a sink''.} Namely, $\PLS$ consists of all search problems that can be reduced (in polynomial-time) to the \emph{Iteration} problem, defined as follows. The input to Iteration consists of two circuits $S,P\colon\{0,1\}^n\to\{0,1\}^n$ (here $S$ is the \emph{successor} circuit and $P$ the \emph{potential} circuit) that implicitly define a dag $G=(V,E)$, $V=\{0,1\}^n$, where $(u,v)\in E$ iff the successor of $u$ is $v$ (i.e., $S(u)=v$) and the potential decreases from $u$ to $v$ (i.e., $P(v)<P(u)$ where $P$'s output is interpreted as an $n$-bit number). The goal is to output the name of any node in $G$ that is a sink (no successor).
\end{example}

In this survey, we have discussed total search problems in the context of more restricted models of computation: proof systems can be viewed as a model of computation solving $\Search(F)$, and communication protocols can solve $\KW(f)$ games. How are these connected to the classical $\TFNP$ theory?

A couple of connections have already been made formal. Given any subclass of $\TFNP$---take $\PLS$ for instance---one may naturally define its query (decision tree) analogue $\PLS^\dt$ and its communication analogue $\PLS^\cc$. For instance, to define $\PLS^\dt$ we start with the query analogue of the Iteration problem, namely, $\Search(F)$ where $F$ is any natural low-width encoding of the contradiction stating that a dag has no sink; for example, one can take as $F$ the famous pebbling formulas. Then $\PLS^\dt$ is defined as consisting of all query search problems that can be reduced, via shallow decision-tree reductions, to the query-Iteration problem. (See~\cite{GKRS2019} for a more formal definition.) What is striking is that in some cases the query analogue of a $\TFNP$ class can be captured by a natural propositional proof system, and the communication analogue can be captured by a natural circuit model (via the KW game). In particular, we have the following classical characterisations discussed in this survey:
\begin{enumerate}
\item[($\FP$)]
The class $\FP$ consists of all search problems soluble in polynomial time on a deterministic Turing machine. By analogy, Karchmer--Wigderson games involve deterministic communication protocols, that is, the communication analogue~$\FP^\cc$ of $\FP$. We may summarise the Karchmer--Wigderson theorem (\autoref{thm:kw-connection}) by saying that \emph{$\FP^\cc$ captures formulas}. Moreover, the deterministic lifting theorem (\autoref{thm:det-lifting}) can be viewed as showing lower bounds for $\FP^\cc$ from its query analogue $\FP^\dt$, which in turn corresponds to the depth of tree-like resolution proofs.

\item[($\PLS$)]
Razborov~\cite{Razborov1995} showed that the circuit complexity of a function $f$ is captured by the least cost of a ``$\PLS$-protocol'' solving the $\KW(f)$ game. Here a \emph{$\PLS$-protocol} is a natural communication analogue of $\PLS$, which turns out to be equivalent to rectangle-dags, which we discussed in \autoref{sec:dags}. We may summarise Razborov's result by saying that \emph{$\PLS^\cc$ captures circuits}. The dag-like lifting theorem \autoref{sec:dags} can then be viewed as showing lower bounds for $\PLS^\cc$ from its query analogue $\PLS^\dt$, which turns out to correspond to the width of resolution proofs.
\end{enumerate}

Both of these connections were found in the 90s. A more recent work has identified a third connection of the above type: $\PPA^\cc$ captures $\mathbb{F}_2$-span programs, and $\PPA^\dt$ captures the $\mathbb{F}_2$-Nullstellensatz proof system~\cite{GKRS2019}. These characterisations, too, could have been discovered in the 90s: The class $\PPA$ (embodying the principle ``every graph with an odd degree vertex has another'') is one of the original $\TFNP$ subclasses defined by Papadimitriou~\cite{Papadimitriou1994}. Similarly, \emph{span programs} are a well-studied circuit model introduced in 1993~\cite{Karchmer1993} with fundamental connections to quantum computing and secret sharing in cryptography.

\begin{figure}[t]
\centering
\begin{tikzpicture}[scale=0.8]
\tikzset{
	inner sep=0,outer sep=3,
	mydash/.style={dashed,color=white!30!black},
	new/.style={line width=1pt,color=myGold},
	marked/.style={color=myPurple},
	model/.style={color=myBlue,font=\itshape}}

\node (a) at (-7.1,0) {\mbox{}};
\node (b) at (7.1,0) {\mbox{}};

\begin{scope}[yscale=1.145]
\node (FP) at (0,0) {$\FP$};
\node (EML) at (0,1.5) {$\EML$};
\node (SML) at (-3,3) {$\SML$};
\node (PPAD) at (3,3) {$\PPAD$};
\node (PPADS) at (0,4.5) {$\PPADS$};
\node (PLS) at (-3,6) {$\PLS$};
\node (PPP) at (0,6) {$\PPP$};
\node (PPA) at (3,6) {$\PPA$};
\node (TFNP) at (0,7.5) {$\TFNP$};
\end{scope}

\node [model, right=0 of FP] {$\displaystyle=\enspace\frac{\text{Formulas}}{\color{myGold}\text{Tree-like Resolution}}$};
\node [model, left=0 of PLS] {$\displaystyle\frac{\text{Circuits}}{\color{myGold}\text{Resolution width}}\enspace =$};
\node [model, right=0 of PPA] {{$\displaystyle=\enspace\frac{\text{$\mathbb{F}_2$-Span programs}}{\color{myGold}\text{$\mathbb{F}_2$-Nullstellensatz}}$}};
\node [model, left=0 of SML] {Comparator circuits\enspace $\leq$};

\path[-{Stealth[length=5pt]},line width=.4pt,black]
(FP) edge (EML)
(EML) edge (SML)
(EML) edge (PPAD)
(SML) edge (PLS)
(SML) edge (PPADS)
(PPAD) edge (PPADS)
(PPAD) edge (PPA)
(PPADS) edge (PPP)
(PLS) edge (TFNP)
(PPP) edge (TFNP)
(PPA) edge (TFNP)
(PPAD) edge[mydash,bend left=31] (PLS)
(PPADS) edge[mydash,bend right=16] (PPA)
(TFNP) edge[mydash,bend right=28] (PPP)
(PLS) edge[mydash,bend right=27] (PPA)
(PPA) edge[mydash,bend right=20] (PPAD)
(EML) edge[mydash,bend right=28] (FP);
\end{tikzpicture}

\vspace{2mm}
\caption{The landscape of $\TFNP$ search problem classes in communication complexity (uncluttered by the usual `$\cc$' superscripts)~\cite{GKRS2019}. A solid arrow $\M_1\rightarrow\M_2$ denotes $\M_1\subseteq\M_2$, and a dashed arrow $\M_1\dashrightarrow\M_2$ denotes $\M_1\nsubseteq\M_2$. Some classes can characterise other models of computation (printed in {\color{myBlue} blue}) while the query complexity analogues are captured by different propositional proof systems (printed in {\color{myGold} yellow}).}
\label{fig:classes}
\vspace{2mm}
\end{figure}
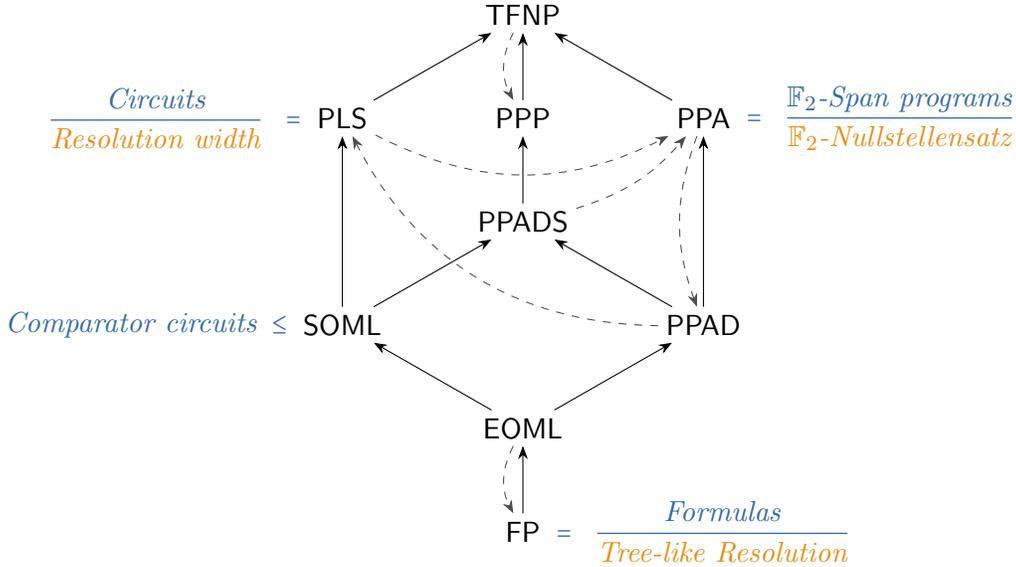

Currently, around a dozen interesting subclasses of $\TFNP$ have been identified. A natural research question, following our philosophy, is to systematically study the interconnections that arise out of the Turing-machine/\-query/\-communication variants of $\TFNP$ subclasses. \autoref{fig:classes} summarises our current understanding of the landscape for communication search problem classes.

\begin{problem}
Complete the picture in \autoref{fig:classes}.
\end{problem}
In particular, the following questions remain open.
\begin{enumerate}[label=(Q\arabic*)]
\item Find circuit models captured by classic classes such as $\PPAD^\cc$, $\PPADS^\cc$, $\PPP^\cc$. What complexity class captures comparator circuits (only an upper bound of $\SML^\cc$ is known)?
\item Find proof systems captured by $\PPAD^\dt$, $\PPADS^\dt$, $\PPP^\dt$.
\item \label{q1}
In query complexity, the relative complexities of classical $\TFNP$ subclasses are nearly completely understood~\cite{Beame1998,BureshOppenheim2004,Morioka2005}. In communication complexity, by contrast, there are huge gaps in our understanding. For example, there are no lower bounds against classes $\PPADS^\cc$ and $\PPP^\cc$ for an explicit search problem.
\item Prove more class separations. If we have two classes, which both admit an associated circuit model, then separating the two classes is often equivalent to separating the two circuit models in the monotone setting. For example, can we show $\SML^\cc\nsubseteq\PPA^\cc$? This is closely related to whether monotone comparator circuits can be more powerful than monotone $\mathbb{F}_2$-span programs (no separation is currently known). 
\item Query-to-communication lifting theorems are known for $\FP$~\cite{Raz1999}, $\PLS$~\cite{GGKS2020}, and $\PPA$~\cite{Pitassi2018}. Prove more. (This is one way to attack \ref{q1} if proved for $\PPADS$ or $\PPP$.)
\end{enumerate}

Building a theory of the intricate relationships between communication search problem classes would further elucidate the relative power of the underlying combinatorial principles and (if available) of the associated circuit models. Moreover, such a theory would likely have applications to all three involved subfields, which is exactly what already happened in the case of the best-understood classes $\FP$, $\PLS$, $\PPA$.

\subsection{Cross-fertilisation of lower-bound methods}

Besides building interconnections between theories, another research direction is to put these connections to use and prove concrete lower-bound results. We propose several problems that can benefit from such cross-fertilisation.

\paragraph{Proof complexity.}
The progress on the central project of proof complexity---proving lower bounds against increasingly powerful proof systems---has been slow. While the limitations of resolution are well understood after decades worth of research, we still lack lower bounds even for \emph{mild generalisations} of resolution. Two of the simplest such generalisations are \emph{resolution over cutting planes}, denoted $\Res(\CP)$~\cite{Krajicek1998}, and \emph{resolution over $\mathbb{F}_2$-linear equations}, denoted $\Res(\lin_2)$~\cite{Raz2008,Itsykson2014}. These are dag-like proof systems that reason using disjunctions of linear inequalities in case of $\Res(\CP)$ or disjunctions of linear equations in case of $\Res(\lin_2)$.%
\begin{problem} \label{ch:res}
Show that $\Res(\CP)$ and $\Res(\lin_2)$ are not polynomially bounded (i.e., there are unsatisfiable formulas that the systems cannot refute with a polynomial-size proof).
\end{problem}

We believe the time is ripe for a breakthrough here. Kraj{\'{\i}}{\v{c}}ek, in his new textbook on proof complexity, writes~\cite[\S22.2]{Krajicek2019}:
\begin{quote}\small
In my view the most pressing open problem is to extend some of the lower bound methods from the A-level to systems like R(CP) or R(LIN). Feasible interpolation seems to be best positioned for that.
\end{quote}
Lifting theorems discussed in this survey have deepened our understanding of the interpolation technique and present a new avenue of attack towards \autoref{ch:res}. In fact, these techniques come frustratingly close to resolving it: Both $\Res(\CP)$ and $\Res(\lin_2)$ involve lines that can be evaluated using a randomized communication protocol. Recent work has already produced lifting theorems both for dag-like proofs whose lines are computed by deterministic protocols~\cite{GGKS2020} and for tree-like proofs whose lines are computed by randomized protocols~\cite{Goos2020bpp}. Seemingly, one only needs to find a way to combine the techniques of these two papers! See also~\cite{Folwarczny2022} for a discussion of this problem.

Another foremost open problem in proof complexity (highlighted in Razborov's survey~\cite{Razborov2016}) concerns semi-algebraic proof systems that manipulate low-degree polynomials, e.g., the ubiquitous \emph{Sum-of-Squares} system. Can we prove lower bounds on their dag-like proof size? Since degree-$d$ polynomials can be efficiently computed by \emph{multi-party number-on-forehead} (NOF) protocols, one might hope to approach this question by developing a lifting theory for NOF protocols. However, our understanding of NOF protocols is lacking even in the tree-like setting.

\paragraph{Circuit complexity.}
An outstanding open problem in circuit complexity concerns matchings:%
\begin{problem} \label{ch:matching}
Show that perfect matching requires exponential-size monotone circuits.
\end{problem}
The best lower bound so far is quasi-polynomial as shown by Razborov~\cite{Razborov1985a}. Matching principles have been studied extensively in proof complexity~\cite{Dantchev2001,Urquhart2003,Alekhnovich2004} and one could hope that these results could be lifted to monotone circuit complexity. There has been an analogous challenge in the study of linear programming formulations of the matching polytope. While there are lifting theorems to prove LP formulation lower bounds~\cite{Chan2016,Kothari2021}, there is no lifting-based proof for Rothvo\ss's~\cite{Rothvos2014} exponential lower bound for the matching polytope. This is one reason why it is still open to show an SDP formulation lower bound for matching: the only lower-bound technique known is a lifting theorem~\cite{Lee2015}. There also remains the lurking prospect that Razborov's lower bound is, in fact, tight! There is a recent cautionary tale this effect: In a surprising breakthrough Dadush and Tiwari~\cite{Dadush2020} showed that Tseitin formulas can be refuted in quasi-polynomial size by the cutting planes proof system, contradicting a conjecture of an exponential lower bound widely believed since the introduction of cutting planes in 1987~\cite{Cook1987}.

\paragraph{Total search problems.}
Finally, in the theory of $\TFNP$, a timely opportunity is to use proof complexity to understand the power of some of the newly introduced $\TFNP$ subclasses; examples include the ``continuous local search'' class $\CLS$~\cite{Daskalakis2011}, ``end of metered line'' class $\EML$~\cite{Hubacek2017} and the ``unique end of potential line'' class~$\UEOPL$~\cite{Fearnley2020}. Last year saw a surprising breakthrough result by Fearnley et al.~\cite{Fearnley2021} who showed that $\CLS$, which was introduced as a natural subclass of the not-so-natural $\PLS\cap\PPAD$, in fact coincides with $\PLS\cap\PPAD$! That such a class collapse was missed for a decade gives an impetus for finding \emph{black-box separations} (aka relativised oracle separations) for the newly introduced classes. Can one show $\CLS^\dt\neq \UEOPL^\dt$, that is, $\CLS\neq\UEOPL$ relative to an oracle? Capturing $\CLS^\dt$ and $\UEOPL^\dt$ in the language of proof complexity is a first step.

\DeclareUrlCommand{\Doi}{\urlstyle{sf}}
\renewcommand{\path}[1]{\small\Doi{#1}}
\renewcommand{\url}[1]{\href{#1}{\small\Doi{#1}}}

\newcommand{\etalchar}[1]{$^{#1}$}

\end{document}